\documentclass[twocolumn,showpacs,amsmath,amssymb]{revtex4}
\usepackage[dvips]{graphics}
\def\imo{i}
\def\re#1{Re(#1)}
\def\im#1{Im(#1)}

\begin{document}
\title{Charged scalar field instability between the event and cosmological horizons}
\author{R. A. Konoplya}\email{konoplya_roma@yahoo.com}
\affiliation{DAMTP, Centre for Mathematical Sciences, University of Cambridge, Wilberforce Road, Cambridge CB3 0WA, United Kingdom.}
\author{Alexander Zhidenko}\email{zhidenko@physik.uni-frankfurt.de}
\affiliation{Institut f\"ur Theoretische Physik, Johann Wolfgang
  Goethe-Universit\"at, Max-von-Laue-Str. 1, 60438 Frankfurt, Germany}
\affiliation{Centro de Matem\'atica, Computa\c{c}\~ao e Cogni\c{c}\~ao,
  Universidade Federal do ABC (UFABC), Rua Aboli\c{c}\~ao, CEP:
  09210-180, Santo Andr\'e, SP, Brazil}
\begin{abstract}
  Recently, a new interesting instability of a charged scalar field in the Reissner-Nordstr\"om-de Sitter background has been found (arXiv:1405.4931v2) through the time-domain integration of the perturbation equation. We investigate further properties of this instability, confirm its existence by concordant frequency-domain and time-domain calculations and show that it occurs at however small value of the coupling $e Q$, where $e$ and $Q$ are charges of a scalar field and black hole respectively. We also investigate the parametric region of instability and show that the critical values of $e Q$ at which the stabilization happens strongly depends on the value of cosmological constant $\Lambda$ and softly on $Q$. We show that all the unstable modes are superradiant, but not all the superradiant modes are unstable. We analytically prove that superradiance is necessary (but not sufficient) condition for the instability in this case and, thereby, demonstrate the superradiant origin of the instability.
\end{abstract}
\pacs{04.30.Nk,04.70.Bw}
\maketitle

\section{Introduction}

Interaction of charged fields with the electromagnetic background of a charged black hole is an active area of research in the contexts of black hole physics \cite{Bekenstein:1973,HodPiran,Hod:2012-2013,Konoplya:2002wt,Li:2012rx,Konoplya:2002ky,Konoplya:2007zx,Li:2013jna,Konoplya:2013rxa,Uchikata:2011zz,Konoplya:2008au,Kokkotas:2010zd} and gauge/gravity duality description of superconductors with high critical temperature \cite{Horowitz:2010nh,Horowitz:2010gk,Gubser:2008px,Son:2013xra,Yao:2013sha,Konoplya:2009hv}. An essential difference of propagation of \emph{charged} fields in the vicinity of charged black holes is that the electromagnetic energy of the black hole can be extracted and carried out by a wave with increased amplitude \cite{Bekenstein:1973}. This effect is called superradiance and is mainly known for rotating black holes \cite{Starobinsky}, for which it allows a wave to carry out part of the black hole's momentum. Frequently, the stability of a black hole or a field in its background cannot be tested analytically, so that a full analysis of quasinormal spectrum of a black hole is necessary (see \cite{Konoplya:2011qq,Ishibashi:2011ws} for recent reviews of black hole stability).  When the field has a nonzero mass, the corresponding bound states (with perfectly reflecting boundary conditions) may be unstable \cite{Herdeiro:2013pia,Degollado:2013bha}, but no real instability occurs upon imposing the correct, \emph{quasinormal} boundary conditions \cite{Konoplya:2013rxa}. Quasinormal modes are proper modes characterizing evolution of perturbation at later times and requiring purely ingoing wave at the event horizon and purely outgoing wave at infinity (or de Sitter horizon).

First study of evolution of perturbation of a charged scalar field in the background of a charged asymptotically flat black hole was performed by S. Hod \cite{HodPiran}, who found asymptotic tails at late times. Further study of the earlier, \emph{quasinormal}, stage of the charged fields perturbation for Reissner-Nordstr\"om \cite{Konoplya:2002wt},  Reissner-Nordstr\"om -anti-de Sitter \cite{Konoplya:2002ky} and Kerr-Newman(-de Sitter) space-times \cite{Konoplya:2007zx,Konoplya:2013rxa} showed that:
\begin{itemize}
  \item Massless conformally coupled charged scalar field and charged Dirac field in the background of Kerr-Newman and Kerr-Newman-de Sitter show no instability in their quasinormal spectrum \cite{Konoplya:2007zx}.
  \item A minimally coupled massive charged scalar field must be stable in the background of an asymptotically flat charged black hole, such as Reissner-Nordstr\"om or Kerr-Newman \cite{Konoplya:2013rxa}.
\end{itemize}

The Reissner-Nordstr\"om-de Sitter is stable against gravitational and electromagnetic perturbations in 4-dimensional space-time, showing the instability only in higher dimensions \cite{Konoplya:2013sba,Konoplya:2014sna}. However, a new interesting instability has been found recently for a minimally coupled charged scalar field in the Reissner-Nordstr\"om-de Sitter background by Z. Zhu and collaborators \cite{Zhu:2014sya}. This instability has a number of peculiar properties:
\begin{enumerate}
  \item It takes place only for the most symmetric $s$-mode of the perturbation, while higher multipoles are stable.
  \item It occurs for small values of $e Q$ coupling, but surprisingly, large values of $e Q$ stabilize the system.
  \item It develops slowly, so that considerable growth is achieved only at times comparable with cosmological scales.
  \item Even relatively small mass of the field acts as a stabilizing factor.
\end{enumerate}

In the present paper we make further investigation of properties of the above instability. First, we confirm the presence of the instability not only by reproducing the time-domain profiles of \cite{Zhu:2014sya}, but also detecting the growing modes in the frequency domain with the help of Frobenius expansion. Next, we analyze the parametric region of instability. Finally, we show that all the growing modes have nonzero real part of $\omega$, which satisfy the condition of superradiance. We prove analytically that the condition of superradiance is necessary but not sufficient for the instability of the charged scalar field in the background of the Riessner-Nordstr\"om-de Sitter space-time.

The paper is organized as follows. Sec. II introduces the basic formula on the background space-time and the equation of motion for a massive charged scalar field. In Sec. III we analyze the connection between superradiance and instability. Sec. IV is devoted to numerical methods used in the paper, while Sec. V is a summary of obtained numerical results.

\section{The basic equations}

The Reissner-Nordstr\"om-de Sitter black hole background can be described by the metric
\begin{eqnarray}
ds^2=-f(r)dt^2+\frac{dr^2}{f(r)}+r^2(d\theta^2+\sin^2\theta d\phi^2)\;,
\end{eqnarray}
where
\begin{eqnarray}
f(r)=1-\frac{2M}{r}+\frac{Q^2}{r^2}-\frac{\Lambda r^2}{3}\,.
\end{eqnarray}
and $M$ and $Q$ are the black hole mass and charge, respectively, $\Lambda$ is the cosmological constant.

We shall designate the Cauchy, event, and cosmological horizons as $r_-$, $r_+$, and $r_c$ respectively ($r_- < r_+ < r_c$).
Then, the metric function $f(r)$ can be written as,
\begin{eqnarray}
f(r)=\frac{\Lambda}{3r^2}(r-r_+)(r-r_-)(r_c-r)(r-r_o)\;.
\end{eqnarray}

A charged, massive scalar field $\psi$ in curved space-time obeys the Klein-Gordon equation
\begin{eqnarray}
[(\nabla^\nu-ieA^\nu)(\nabla_\nu-ieA_\nu)-\mu^2]\psi=0\;,
\end{eqnarray}
where $e$ and $\mu$ are, respectively, the charge and mass of
the field and $A_\mu=-\delta^0_\mu Q/r$ is the electromagnetic 4-potential of the black hole.

After the standard separation of angular variables with the help of spherical harmonics and introduction a new wave function $\Psi$, the above equation of motion can be reduced to the following form:
\begin{eqnarray}
-\frac{\partial^2\Psi}{\partial t^2}+\frac{\partial^2\Psi}{\partial r_*^2}-2i\Phi\frac{\partial\Psi}{\partial t}+(\Phi^2-V)\Psi=0\;,\label{radial_eq}
\end{eqnarray}
where $dr_*=\frac{dr}{f(r)}$ is the tortoise coordinate, and
\begin{eqnarray}
\Phi(r)=\frac{eQ}{r},\quad V(r)=f(r)\bigg(\mu^2+\frac{l(l+1)}{r^2}+\frac{f'(r)}{r}\bigg)\;.\label{V}
\end{eqnarray}

\section{Superradiance and instability}

In order to analyze superradiance around the Reissner-Nordstr\"om-de Sitter black hole, we consider the scattering problem for a charged scalar field.
Owing to the $\partial_t$ Killing vector for stationary space-times, the following ansatz $\Psi(r,t)=\exp[-\imo\omega t]\Psi(r)$ transforms (\ref{radial_eq}) to the Sch\"ordinger-like form
\begin{eqnarray}
{\frac{d^2\Psi}{dr_*^2}}+(\omega-\Phi(r))^2\Psi-V(r)\Psi=0\;.\label{radial}
\end{eqnarray}
Due to the $\re{\omega}\rightarrow-\re{\omega}$, $\Phi(r)\rightarrow-\Phi(r)$ symmetry of (\ref{radial}), consideration of the $eQ>0$ case only is sufficient.

We shall be using the standard scattering boundary conditions, implying that the wave coming from the cosmological horizon will be partially reflected from the potential barrier and come back to the cosmological horizon, while at the event horizon purely in-coming wave is always required:
\begin{eqnarray}
\Psi\sim\left\{
  \begin{array}{ll}
    Be^{-i(\omega-\Phi(r_+))r_*}, & r\rightarrow r_+, \\
    e^{-i(\omega-\Phi(r_c))r_*}+A e^{+i(\omega-\Phi(r_c))r_*}, & r\rightarrow r_c.
  \end{array}
\right.
\end{eqnarray}
Since the Wronskian of the complex conjugated solutions is constant, one can derive the following relation
\begin{eqnarray}
|A|^2=1-{\frac{\omega-\Phi(r_+)}{\omega-\Phi(r_c)}}|B|^2\;.
\end{eqnarray}
Amplification of the incident wave, that is, a \emph{superradiance}, occurs when
\begin{eqnarray}
\Phi(r_c)=\frac{eQ}{r_c}<\omega<\frac{eQ}{r_+}=\Phi(r_+).\label{regime}
\end{eqnarray}
When the cosmological constant $\Lambda$ approaches zero, the above superradiant condition Eq.~(\ref{regime}) is reduced to the one for asymptotically flat space-times \cite{Bekenstein:1973,Hod:2012-2013} ($\omega$ is the real oscillation frequency).

Let us now understand how the condition of superradiance is related to instability, that is, to the possible presence of growing quasinormal modes in the black hole's spectrum. For this, let us consider the quasinormal (and not scattering) boundary conditions and imply that frequencies, generally, are complex:

\begin{eqnarray}\label{QNBC}
\Psi\sim\left\{
  \begin{array}{rcl}
    e^{-i(\omega-\Phi(r_+))r_*}, \quad & r\rightarrow r_+, &\quad (r_* \rightarrow -\infty), \\
    e^{+i(\omega-\Phi(r_c))r_*}, \quad & r\rightarrow r_c, &\quad (r_* \rightarrow +\infty).
  \end{array}
\right.
\end{eqnarray}

We shall prove that the real part of $\omega$ satisfying (\ref{regime}) is the necessary, but not sufficient, condition for the instability.

First, let us multiply (\ref{radial}) by the complex conjugated $\Psi^\star$ and integrate the first term by parts,
\begin{eqnarray}
\Psi^\star(r_\star)\Psi^\prime(r_\star)\Biggr|_{-\infty}^\infty + \intop_{-\infty}^\infty (\omega-\Phi)^2|\Psi(r_\star)|^2dr_\star \\\nonumber=\intop_{-\infty}^\infty\left(V|\Psi(r_\star)|^2 + |\Psi^\prime(r_\star)|^2\right)dr_\star.
\end{eqnarray}
The righthand side is real and positive since the effective potential $V$ is positive. Because $\Phi(r)$ is a monotonically decreasing function, taking imaginary part of both sides, we find that if either $\re{\omega}\geq\Phi(r_+)\geq\Phi(r)$ or $\re{\omega}\leq\Phi(r_c)\leq\Phi(r)$ is satisfied, then we have $\im{\omega}<0$. This proves that the instability can take place only if
\begin{eqnarray}\label{superradiant}
\Phi(r_c)=\frac{eQ}{r_c}<\re{\omega}<\frac{eQ}{r_+}=\Phi(r_+).
\end{eqnarray}

\begin{figure}
\resizebox{\linewidth}{!}{\includegraphics*{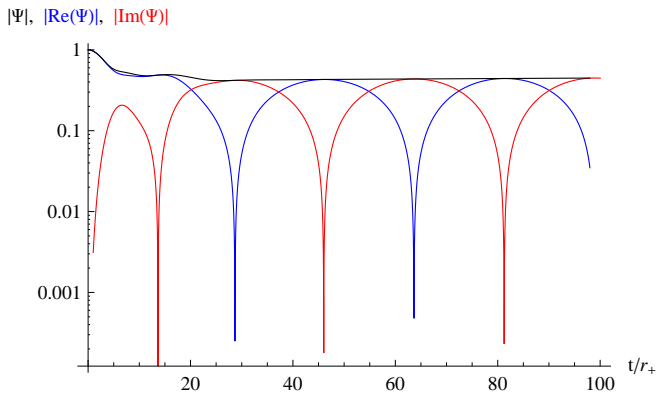}}
\caption{Real and imaginary part of the signal for $eQ=0.4$, $r_-=0.07r_+$, $r_+=0.186r_c$ ($Q=0.5M$, $\Lambda M^2=0.023$).
The dominant frequency is $\omega r_+ = 0.089160+0.000754\imo$.}\label{fig5}
\end{figure}

Indeed, in the time-domain profiles (fig. \ref{fig5}) we observe that the real part for all the unstable modes we found satisfy (\ref{superradiant}). For $eQ$ larger than the threshold of instability the real part of the dominant mode also satisfies (\ref{superradiant}), being stable. This is qualitatively different from the higher-dimensional asymptotically anti-de Sitter black holes, for which the necessary condition is also the sufficient one \cite{Kodama:2009rq,Wang:2014eha}.

\section{Time-domain and frequency-domain analysis}

Here in order to integrate the wave equation (\ref{radial_eq}) and analyze the spectrum of the perturbation, we shall use, first of all, the time-domain integration, which includes contribution from all modes.
The discretization scheme which we shall use was proposed in \cite{Abdalla:2010}.
Defining $\Psi(r_*,t)=\Psi(j\Delta r_*, i\Delta t)=\Psi_{j,i}$, $V(r(r_*))=V(j\Delta r_*)=V_j$ and $\Phi(r(r_*))=\Phi (j\Delta r_*)=\Phi_j$, we can write down (\ref{radial_eq}) as
\begin{eqnarray}
&&-\frac{(\Psi_{j,i+1}-2\Psi_{j,i}+\Psi_{j,i-1})}{\Delta
t^2}-2i\Phi_j\frac{(\Psi_{j,i+1}-\Psi_{j,i-1})}{2\Delta t} \nonumber \\ \nonumber
&&+\frac{(\Psi_{j+1,i}-2\Psi_{j,i}+\Psi_{j-1,i})}{\Delta
r_*^2}-V_j\Psi_{j,i}={\cal O}(\Delta t,\Delta
r_*).
\end{eqnarray}
The initial Gaussian wave-package has the form
$\Psi(r_*,t=0)=\exp\bigg[-{(r_*-a)^2/2b^2}\bigg]$, $\Psi(r_*,t<0)=0$.
Then, the evolution of $\Psi$ can be described by the following expression
\begin{eqnarray}
\Psi_{j,i+1}=-{\frac{(1-i\Phi_j\Delta t)\Psi_{j,i-1}}{1+i\Phi_j\Delta t}}+{\frac{\Delta t^2}{\Delta r_*^2}}
{\frac{\Psi_{j+1,i}+\Psi_{j-1,i}}{1+i\Phi_j\Delta t}} \nonumber\\\nonumber
+\bigg(2-2{\frac{\Delta t^2}{\Delta r_*^2}}-\Delta t^2V_j\bigg){\frac{\Psi_{j,i}}{1+i\Phi_j\Delta t}}\;.
\end{eqnarray}
Following \cite{Abdalla:2010}, we choose the parameters $a=0$ and $b=\sqrt{10}$ in the Gaussian wave package and use $${\frac{\Delta t}{\Delta r_*}}=\frac{1}{2}<1,$$ making sure that $\Delta t$ is small enough to achieve the required precision of the profile.

Real part of $\omega$ cannot be extracted from a time-domain profile obtained with the help of the above integration for $|\Psi(r=const,t)|$ because charged field perturbations are complex. In \cite{Zhu:2014sya} it was erroneously reported that the growing modes have zero real part and, therefore, cannot be superradiant.
It turns out, that while absolute value of $\Psi$ does not oscillate at the late times, its phase continues changing, implying nonzero real part of $\omega$. In order to calculate $\re{\omega}$ we used the Prony method of fitting the time-domain profile data by superposition of damping exponents \cite{Berti:2007dg}
\begin{equation}\label{damping-exponents}
\Psi(r,t)\simeq\sum_{i=1}^pC_ie^{-\imo\omega_i (t-t_0)}.
\end{equation}
We consider a late time period, which starts at $t_0$ and ends at $t=N\Delta t+t_0$, where $N$ is an integer and $N\geq2p-1$. Then the formula (\ref{damping-exponents}) is valid for each value from the profile data:
\begin{equation}
x_n\equiv\Psi(r,n\Delta t+t_0)=\sum_{j=1}^pC_je^{-\imo\omega_j n\Delta t}=\sum_{j=1}^pC_jz_j^n.
\end{equation}
The Prony method allows us to find $z_i$ in terms of the known $x_n$ and, since $\Delta t$ is also known, to calculate the quasinormal frequencies $\omega_i$.

In the frequency domain we shall use the well-known method of Frobenius expansion \cite{Leaver:1985ax}. We express $\Psi$ in (\ref{radial}) as a product of the series, convergent in $r_+\leq r<r_c$, and a singular factor, which is fixed by the quasinormal boundary conditions (\ref{QNBC}) in two regular singular points, $r_+$ and $r_c$. This choice implies that the series are convergent in $r=r_c$ if and only if the corresponding solution to (\ref{radial}) satisfies the quasinormal boundary conditions. One can also find the recurrence relation for the series coefficients, which in our case has 7 terms. Using the Gaussian elimination we reduce the order of the recurrence relation. Finally, from the condition for the series convergence at $r=r_c$ we find an equation with the infinite continued fraction, which can be solved numerically with respect to the quasinormal frequency $\omega$. For more details we address the reader to our earlier paper \cite{Konoplya:2007zx} where we considered the conformally coupled charged scalar field in the same background.

\section{Numerical results}

\begin{figure}
\resizebox{\linewidth}{!}{\includegraphics*{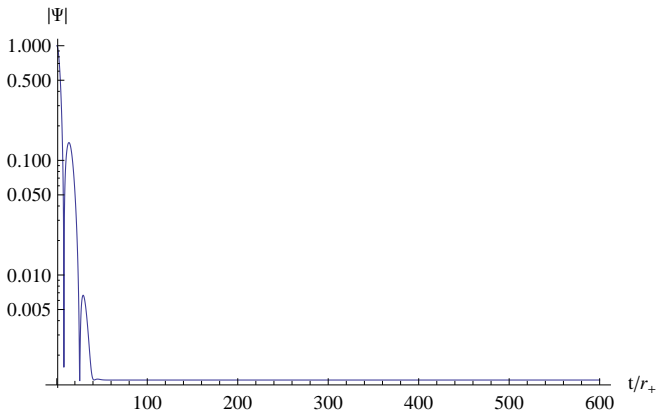}}
\caption{Time-domain profile for $eQ=0.001$, $\mu=0$, $r_-=0$, $r_+=0.2r_c$ ($Q \rightarrow 0$, $\Lambda M^2\approx0.023$).}\label{fig1}
\end{figure}

The question which could be addressed first is for which values of $e Q$ and $\Lambda$ the instability takes places. We conclude that the onset of instability occurs for \emph{arbitrarily small values of} $e Q$ and $\Lambda$. An illustration of this can be seen on fig.~\ref{fig1}, where the instability with very small growth rate $e Q=0.001$ is shown. For smaller $e Q$ the unstable mode has even smaller growth rate, which vanishes when one goes to the limit $e Q \rightarrow 0$. For the exact value $e Q =0$, there is no static mode and the normal fall-off takes place instead. For neutral fields the quasinormal modes of asymptotically de Sitter black holes approach quasinormal modes of asymptotically flat black holes in the limit $\Lambda \rightarrow 0$ \cite{Konoplya:2004uk}. As we can see here, for a charged scalar field this is not true anymore, because a new unstable mode appear for however small $\Lambda$.

\begin{table}
\caption{Crossing the threshold of instability for $r_-=0.07r_+$, $r_+=0.186r_c$ ($Q\approx0.5M$, $\Lambda M^2\approx0.023$).}\label{tabl1}
\begin{tabular}{|c|c|c|}
\hline
$eQ$&$\omega r_+$&$eQr_+/r_c$\\
\hline
$0.40$&$0.08916+0.00075\imo$&$0.07440$\\
$0.41$&$0.09149+0.00067\imo$&$0.07626$\\
$0.42$&$0.09383+0.00058\imo$&$0.07812$\\
$0.43$&$0.09617+0.00048\imo$&$0.07998$\\
$0.44$&$0.09850+0.00037\imo$&$0.08184$\\
$0.45$&$0.10083+0.00023\imo$&$0.08370$\\
$0.46$&$0.10317+0.00011\imo$&$0.08556$\\
$0.47$&$0.10551-0.00002\imo$&$0.08742$\\
$0.48$&$0.10784-0.00017\imo$&$0.08928$\\
$0.49$&$0.11017-0.00033\imo$&$0.09114$\\
\hline
\end{tabular}
\end{table}

\begin{figure}
\resizebox{\linewidth}{!}{\includegraphics*{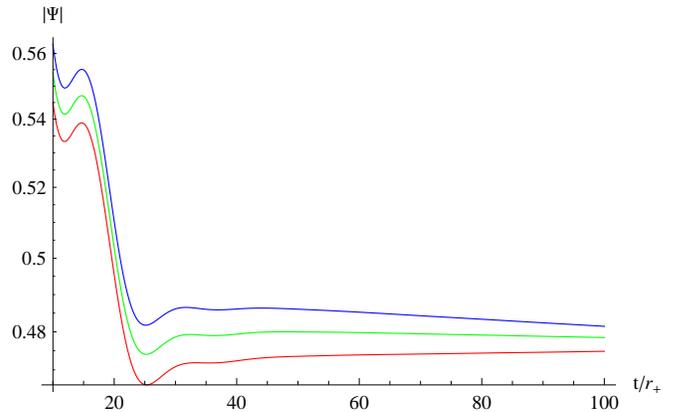}}
\caption{Time-domain profiles for $eQ=0.43$ (red, growing) $eQ=0.44$ (green) $eQ=0.45$ (blue, decaying), $\mu=0$, $r_-=0$, $r_+=0.2r_c$
($Q \rightarrow 0$, $\Lambda M^2\approx0.023$).}\label{fig2}
\end{figure}

\begin{figure}
\resizebox{\linewidth}{!}{\includegraphics*{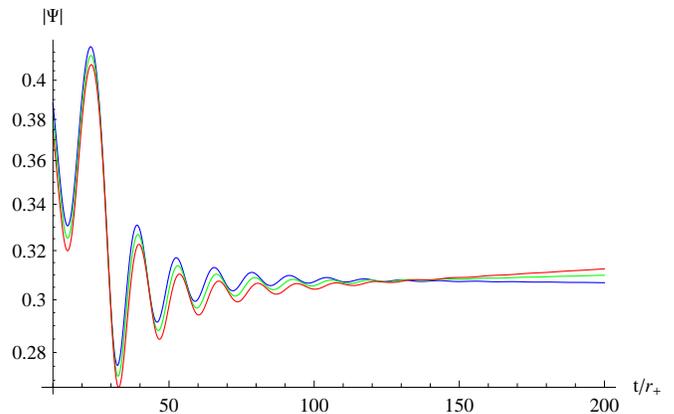}}
\caption{Time-domain profiles for $eQ=0.51$ (red, growing) $eQ=0.52$ (green) $eQ=0.53$ (blue, decaying), $\mu=0$, $r_-=0.837r_+$, $r_+=0.106r_c$ ($Q\approx M$, $\Lambda M^2\approx0.023$).}\label{fig3}
\end{figure}

On figs.~\ref{fig2}~and~\ref{fig3} one can see the time-domain profiles for various values of $e Q$ near the onset of instability: for
almost neutral black hole $Q \rightarrow 0$, but nonzero $e Q$ (fig.~\ref{fig2}) and for highly charged black hole $Q \approx M$ (fig.~\ref{fig3}). Comparison of these two plots shows that the black hole charge $Q$ only softly increases the threshold value of $e Q$ for the onset of instability.
On the contrary, larger cosmological constant $\Lambda$ significantly diminishes the region of instability as can be seen from fig.~\ref{instreg}.

\begin{table}
\caption{Dominant modes for $r_-\approx0.0717r_+$, $r_+\approx0.0108r_c$ ($Q\approx0.50M$, $\Lambda M^2\approx0.0001$).}\label{tabl2}
\begin{tabular}{|c|c|c|c|}
\hline
$eQ$&$\omega r_+$ (time domain)&$\omega r_+$ (frequency domain)&$eQr_+/r_c$\\
\hline
$0.10$&$0.001099\!\!+\!0.000044\imo$&$0.00109917\!\!+\!0.000044298\imo$&$0.00108$\\
$0.50$&$0.00638+0.00104\imo$&$0.00637557+0.00102996\imo$&$0.00541$\\
$0.75$&$0.01080+0.00049\imo$&$0.0107901+0.000490746\imo$&$0.00813$\\
$0.80$&$0.01150+0.00023\imo$&$0.0115066+0.000224642\imo$&$0.00867$\\
$0.85$&$0.01215-0.00002\imo$&$0.0121627-0.000025554\imo$&$0.00921$\\
$0.90$&$0.01277-0.00024\imo$&$0.0127799-0.000248131\imo$&$0.00975$\\
\hline
\end{tabular}
\end{table}

From Table I and II we can see that all the growing modes satisfy the superradiance condition (\ref{regime}), which is necessary but not sufficient condition for the instability. That is why, some decayed modes in Table I and II in the stable sector also satisfy the superradiance condition (\ref{regime}). The fact that all growing modes satisfy (\ref{regime}) is a strong indication that the found instability is of superradiant nature.
From Table II we can also see that the time-domain and frequency-domain calculations are in excellent concordance. Note, that in \cite{Konoplya:2007zx} no instability for the massless charged scalar field in the Kerr-Newman black hole was observed, because the scalar field was taken in its conformal form (despite eq.~(4) in \cite{Konoplya:2007zx} is written for the minimally coupled field).

\begin{figure}
\resizebox{\linewidth}{!}{\includegraphics*{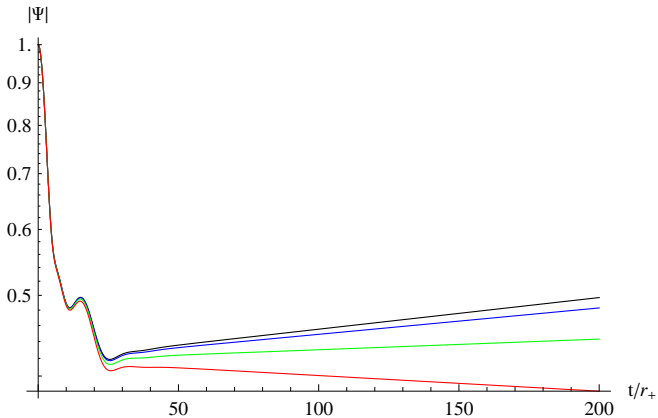}}
\caption{Time-domain profiles for charged $eQ=0.4$ massless (black) and massive: $\mu r_+=0.01$ (blue), $\mu r_+=0.02$ (green), and $\mu r_+=0.03$ (red, stable) scalar field in Reissner-Nordstr\"om-de Sitter background: $r_-=0.07r_+$, $r_+=0.186r_c$ ($Q=0.5M$, $\Lambda M^2=0.023$).}\label{fig4}
\end{figure}

The smaller $\Lambda$, the smaller is the instability growth rate, so that in the real world the found instability should have so small growth rate
that it should practically look like a kind of scalar ``hair''. The exponential growth could be significant only at cosmological times. Moreover, even a very small but nonzero mass of the field $\mu$ would stabilize the scalar field, though for large $\Lambda$ one needs much larger $\mu$ for stabilization. In addition, the instability requires $e Q \gg \mu M$ which means that the electromagnetic repulsion must much be larger than the gravitational attraction of a charged particle to the black hole. Thus, the instability is far beyond the accretion limit for a charged black hole (see fig.~\ref{fig4}). Fig.~\ref{instreg} shows that for small $\Lambda$, the critical value of $e Q$ at which the stabilization occurs quickly diminishes. This certainly does not contradict to the fact that the instability takes place for however small $\Lambda$.

\begin{figure}
\resizebox{\linewidth}{!}{\includegraphics*{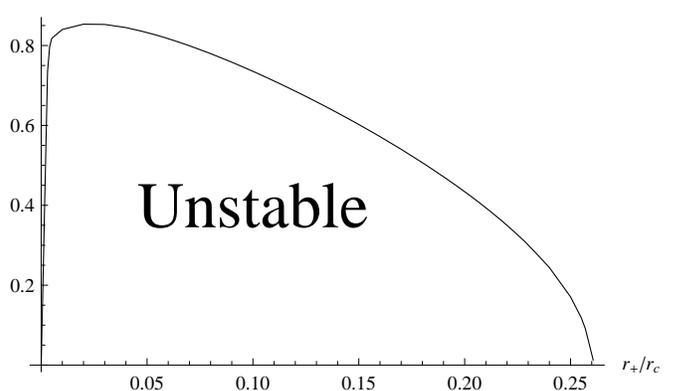}}
\caption{The instability region for a weakly charged Reissner-Nordstr\"om-de Sitter black hole ($Q \rightarrow 0$, $eQ\neq0$).}\label{instreg}
\end{figure}

\section{Conclusions}

The main result of this work is \emph{observation of a superradiant nature of the recently found instability} of a charged scalar field in the background of the 4-dimensional Reissner-Nordstr\"om-de Sitter black hole. It is evident that the distribution of the extracted from the black hole electromagnetic energy between the event horizon and cosmological horizon leads to such a bizarre instability. A further interesting step in this direction could be a nonlinear analysis of collapse of a charged field in the de Sitter space.

The instability takes place for the spherically symmetric perturbation $\ell =0$, while the superradiant condition is hold also for higher $\ell$. Thus, once the instability has superradiant origin, it would be natural to suspect that there are growing modes for some range of parameters also at higher multipoles. We tested a few first $\ell$ at moderately large values of $e Q$ and found no unstable modes, though one should not exclude this possibility \emph{ad hoc}. Apparently, higher multipole moment $\ell$ increases the height of the potential barrier and makes the field more stable.

The stability of higher multipoles is closely related also to the perturbations of charged nonzero spin fields in the Reissner-Nordstr\"om background.   As the first dynamical mode of the charged Dirac field  starts not from $\ell =0$, but effectively from $\ell = 1/2$, while for higher spins the lowest value of $\ell$ is larger, one should not expect any instability of this kind for charged fields with nonzero spin. Indeed, careful search in the frequency domain \cite{Konoplya:2007zx} for a charged Dirac field shows no growing modes in the spectrum. However, an additional check with the help of  the time-domain integration would guarantee us that no unstable mode was lost in the frequency-domain search.

\begin{acknowledgments}
A.~Z. was supported by the Alexander von Humboldt Foundation, Germany and Coordena\c{c}\~ao de Aperfei\c{c}oamento de Pessoal de N\'ivel Superior (CAPES), Brazil.
\end{acknowledgments}

\end{document}